\documentclass{article}
\usepackage{iclr2026_conference,times}
\usepackage[utf8]{inputenc}

\usepackage{booktabs}
\usepackage{multirow}
\usepackage{threeparttable}
\usepackage{graphicx}
\usepackage{float}
\usepackage{caption}
\usepackage{tabularx}
\usepackage{makecell}
\usepackage{ragged2e}
\usepackage{algorithm}
\usepackage{algpseudocode}
\usepackage{adjustbox}
\usepackage{array}
\usepackage{xcolor}

\definecolor{lvl0bg}{RGB}{50, 60, 70}
\definecolor{lvl0fg}{RGB}{255, 255, 255}
\definecolor{lvl3bg}{RGB}{238, 242, 246}
\definecolor{catavgbg}{RGB}{220, 228, 236}

\newcolumntype{D}{>{\raggedright\arraybackslash}X}
\newcolumntype{N}{>{\centering\arraybackslash}X}
\newcolumntype{C}[1]{>{\centering\arraybackslash}m{#1}}

\usepackage{amsmath,amsfonts,bm}

\def\eqref#1{equation~\ref{#1}}

\def\1{\bm{1}}

\DeclareMathAlphabet{\mathsfit}{\encodingdefault}{\sfdefault}{m}{sl}
\SetMathAlphabet{\mathsfit}{bold}{\encodingdefault}{\sfdefault}{bx}{n}

\usepackage{hyperref}
\usepackage{url}

\title{ProcCtrlBench: Evaluating Process-Level Defects and Control Preservation in LLM Coding Agents}

\author{
Jiawei He, Jie Jia, Chenbo Liu, Chaoyi Xue, Yapeng Song, Xikai Yang$^{*}$, Dong Sun$^{*}$ \\
Amap, Alibaba Group \\
$^{*}$Corresponding authors
}

\iclrfinalcopy

\begin{document}

\maketitle

\begin{abstract}
Existing benchmarks for LLM coding agents primarily evaluate final outcomes. 
While useful for measuring overall capability, these metrics provide limited visibility and often miss defects that arise during execution.
We present ProcCtrlBench, a benchmark for execution-process evaluation in LLM coding agents. 
ProcCtrlBench organizes recurrent execution defects into a reusable ontology covering 11 defect types in 4 categories, and evaluates agent trajectories through standardized process evidence rather than final outcomes alone. 
To support comparison across heterogeneous agents, ProcCtrlBench standardizes raw logs into a unified trajectory representation and reports calibrated scorecards over process-level findings. 
In addition, ProcCtrlBench uses control preservation as a way to quantify execution-process quality, capturing whether execution remains interpretable, interruptible, correctable, reversible, and able to hand back authority when needed. 
We evaluate ProcCtrlBench on 200 cases sampled from three benchmarks: AndroidBench, TerminalBench, and SWE-bench-Verified. 
Results show that ProcCtrlBench can be instantiated with useful reliability, provides more stable semantics than direct thresholding, and reveals meaningful differences in execution quality that are often overlooked by conventional outcome-based evaluation.

\end{abstract}

\section{Introduction}

Existing benchmarks for LLM coding agents primarily evaluate final outcomes, such as task completion, compilation success, and test pass rates \citep{jimenez2024swe,merrill2026terminal,chen2021evaluating}. 
These metrics \citep{bean2026measuring,wallach2025position} are useful for measuring overall capability, but they provide limited visibility into how an agent executes a multi-step task. 
In realistic coding settings, many important defects arise during execution rather than only at the final state. An agent may eventually solve a task while following a trajectory that is repetitive, inefficient, weakly structured, or difficult to supervise \citep{kwa2025measuring}. As coding agents become more autonomous, this gap between final-outcome evaluation and execution-process quality becomes increasingly important \citep{schwartz2026circle}.

Many practically important failures arise during execution rather than only at the endpoint \citep{salaudeen2025measurement,ibrahim2025towards}. 
As a result, two runs with similar final outcomes may still differ substantially in execution quality, diagnosability, and practical usability \citep{liao2024ai}. 
This limitation of endpoint-based evaluation is increasingly recognized beyond coding agents. ProcBench \citep{fujisawa2024procbench}, for example, directly evaluates multi-step inference instead of relying only on final-answer accuracy. TRACE \citep{wang2023trace} likewise shows that, under continual learning, preserving reasoning-related behavior is important because adaptation can degrade general capability and instruction following. 
Together, these works support our view that coding-agent evaluation should consider not only whether a task is solved, but also how reliably the execution process is maintained.

To address this limitation \citep{chouldechova2024shared,weidinger2025toward}, we present ProcCtrlBench, a benchmark for execution-process evaluation in LLM coding agents. ProcCtrlBench organizes recurrent execution defects into a reusable ontology covering 11 defect types in four categories: context management, tool-use efficiency, workflow architecture, and tool-ecosystem consistency. Rather than relying on final outcomes alone, ProcCtrlBench evaluates complete agent trajectories through standardized process evidence. In addition, ProcCtrlBench uses control preservation as a way to quantify execution-process quality, capturing whether execution remains interpretable, interruptible, correctable, reversible, and able to hand back authority when needed. The goal is not to replace outcome-based evaluation, but to complement it with a structured benchmark of execution quality.

To support comparison across heterogeneous agents, ProcCtrlBench standardizes raw logs into a unified trajectory representation and reports calibrated scorecards over process-level findings. The benchmark operates on complete execution trajectories rather than final outputs alone. It first standardizes heterogeneous logs into a unified event representation, then extracts defect-specific evidence and reports calibrated assessments of process-level findings. The resulting scorecard includes defect-level findings, dimension-level process-quality scores, control-preservation scores, and a compact summary score for coarse comparison.

We evaluate ProcCtrlBench on 200 cases sampled from three benchmarks: AndroidBench, TerminalBench \citep{merrill2026terminal}, and SWE-bench-Verified \citep{jimenez2024swe}. Our results show that ProcCtrlBench can be instantiated with useful reliability, provides more stable semantics than direct thresholding, and reveals meaningful differences in execution quality that are often overlooked by conventional outcome-based evaluation.

Our contributions are summarized as follows:
\begin{itemize}
    \item We propose ProcCtrlBench, a benchmark for execution-process evaluation in LLM coding agents beyond final outcomes.
    \item We introduce a reusable ontology of execution defects, covering 11 defect types in four categories, to support structured benchmarking of agent trajectories.
    \item We standardize heterogeneous execution logs into a unified trajectory representation and produce calibrated scorecards over process-level findings, with control preservation used to quantify execution-process quality.
    \item We evaluate ProcCtrlBench on 200 cases sampled from AndroidBench, TerminalBench, and SWE-bench-Verified, and show that it can be instantiated with useful reliability and surfaces meaningful system differences beyond outcome-based evaluation.
\end{itemize}

\section{Related Work}

Research on LLM evaluation has largely emphasized outcome-oriented benchmarks \citep{kiela2021dynabench,srivastava2023beyond,eriksson2025can,bean2026measuring}. 
These benchmarks are useful for measuring capability on standardized tasks, but they mostly assess whether a system succeeds at the end rather than how it reaches that result \citep{schwartz2026circle}. 
This limitation is especially pronounced for multi-step agents, where important failures often arise during execution, such as context degradation, repetitive tool use, inefficient action chains, and loss of control \citep{cemri2026multi,weidinger2023sociotechnical}.
As a result, recent work has increasingly moved toward process-sensitive evaluation. 
For example, ProcBench \citep{fujisawa2024procbench} evaluates multi-step inference by testing whether models can correctly follow explicit procedures, and TRACE \citep{wang2023trace} shows that preserving reasoning-related behavior is important for maintaining capability under continual learning. 
These efforts highlight the value of going beyond endpoint success. In this spirit, ProcCtrlBench focuses on execution-process defects and control preservation in coding agents.

Accordingly, recent work \citep{liu2024agentbench,zhu2026establishing} has increasingly examined intermediate agent behavior, including planning traces, tool-use trajectories, error recovery, and failure analysis in realistic environments. 
These studies \citep{weidinger2023sociotechnical,liao2024ai,schwartz2026circle} show that the execution process is central to practical agent reliability and that benchmark success does not necessarily imply robust behavior. 
Still, much of this work remains focused on qualitative diagnosis, localized behavioral analysis, or task-specific case studies \citep{ibrahim2025towards,berman2024scoping}. 
Our goal differs in scope: we seek to define a reusable benchmark space that organizes recurring process failures into comparable benchmark points, thereby supporting reusable benchmark construction and structured cross-system comparison.

Some related research \citep{guo2017calibration,sullivanintroduction,he2026survey,shorinwa2025survey,huang2024survey} focuses on uncertainty estimation and calibration for machine learning systems. 
Methods such as confidence estimation \citep{detommaso2024multicalibration,lin2024generating}, disagreement-based approaches \citep{xiong2024can}, and post hoc calibration \citep{ulmer2024calibrating} improve the interpretability of model outputs, but they are typically applied to single-turn responses or final predictions. 
ProcCtrlBench instead applies calibration to process-level findings derived from sequential execution trajectories. The aim is not simply to estimate confidence in an output, but to assign stable, comparable semantics to benchmark findings across defect types and task settings.

Work on human oversight and control in autonomous systems is also closely related \citep{weidinger2023sociotechnical,flournoy2020building,manheim2025limits}. This literature emphasizes that reliable automation depends not only on task success, but also on whether system behavior remains understandable, interruptible, correctable, and safe to hand back \citep{matias2023humans}. 
These concerns are particularly important for long-horizon coding agents, where a seemingly successful run may still be opaque, brittle, or difficult to supervise \citep{lin2026scaling,schwartz2026circle}. 
ProcCtrlBench extends this perspective by using control preservation to quantify execution-process quality and integrating it with process-defect findings in a unified scorecard.

\section{Methods}

\subsection{ProcCtrlBench Overview}

\begin{figure*}[t]
\centering
\includegraphics[width=0.97\textwidth]{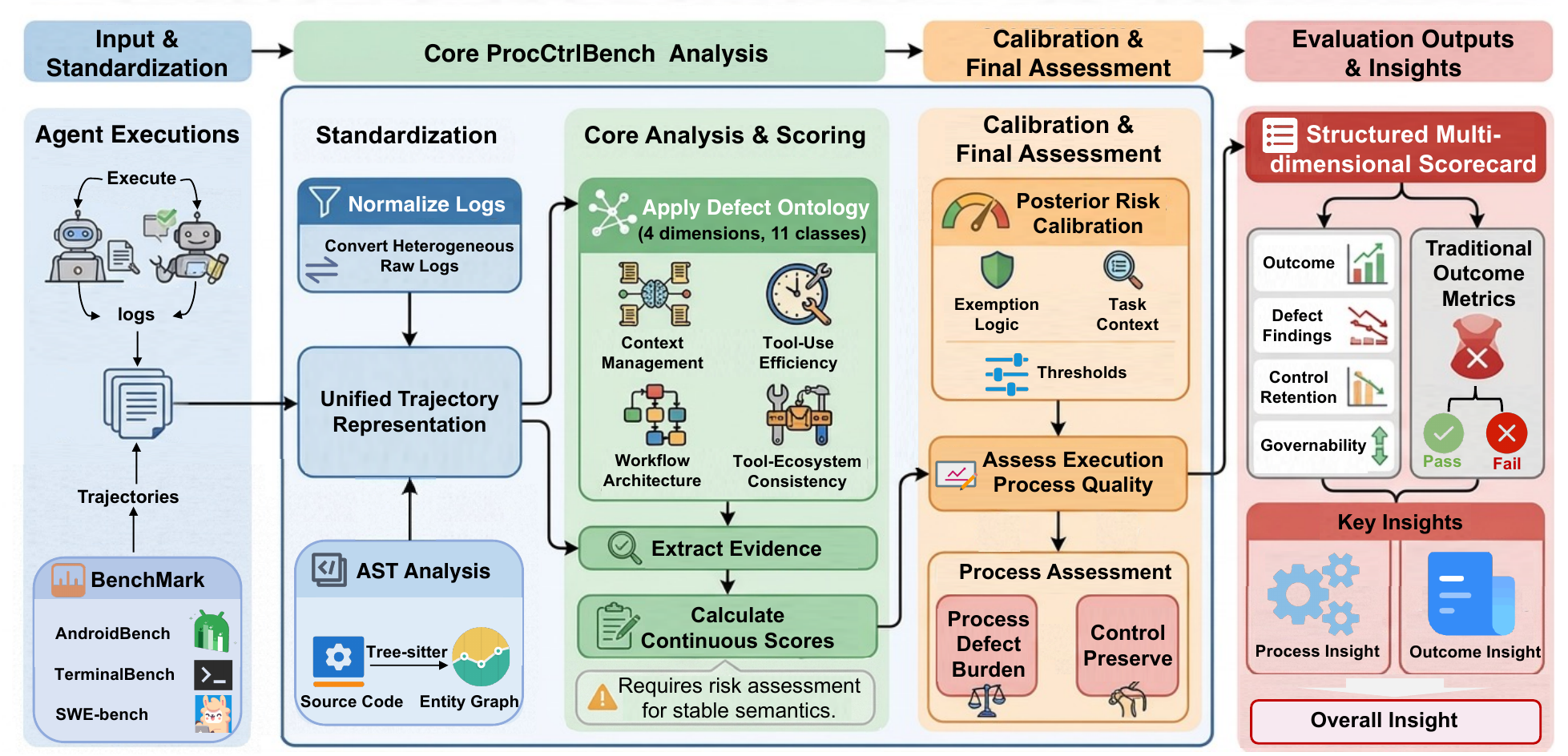}
\caption{Overview of ProcCtrlBench. ProcCtrlBench transforms heterogeneous coding-agent trajectories into calibrated process scorecards for execution-process evaluation.}
\label{fig:procbench_overview}
\end{figure*}

Figure~\ref{fig:procbench_overview} summarizes the overall design of ProcCtrlBench. 
The central challenge is that different coding agents expose heterogeneous execution logs, tool traces, and workflow structures, which makes direct process comparison difficult. \citep{wang2025openhands,zhu2026establishing}. 
ProcCtrlBench addresses this by first transforming raw executions into a standardized trajectory representation, so that different systems can be evaluated within a common process space.

On top of this unified representation, ProcCtrlBench applies a reusable ontology of execution defects covering 11 defect types in four categories.
Rather than treating each detector output as a final label, ProcCtrlBench separates three stages: evidence extraction, risk estimation, and scorecard reporting. 
This separation is important because many process defects are matters of degree and context rather than naturally binary events.

Raw detector outputs are then mapped to calibrated posterior risks under task context and exemption logic. 
The purpose of calibration is to improve comparability and interpretability across defect types and task settings. 
Finally, ProcCtrlBench aggregates defect findings into a structured scorecard that reports process-defect burden, dimension-level quality scores, control preservation, and a compact overall summary. 
In our presentation, the full scorecard is the primary output; the scalar summary is intended only for coarse comparison.

\subsection{Standardized Trajectory Representation}

ProcCtrlBench compares heterogeneous agent systems through a standardized trajectory representation rather than raw platform-specific logs. 
Coding agents differ in logging formats, tool interfaces, message schemas, and workflow structures; without standardization, process-level findings would be difficult to compare across systems.

Given a raw execution log \(L\), we apply a deterministic mapping
\[
\Psi : L \rightarrow T,
\]
where
\[
T = \{e_1, e_2, \ldots, e_n\}
\]
is an ordered sequence of standardized events. Each event is represented as
\[
e_i = (t_i, m_i, u_i, v_i, o_i, c_i, d_i),
\]
where \(t_i\) denotes event type, \(m_i\) the textual payload, \(u_i\) the tool invocation, \(v_i\) the tool return or validation result, \(o_i\) the observable external operation, \(c_i\) the context state, and \(d_i\) the structural dependency information, including the parent step, branch identity, workflow unit, or agent identity.

This layer is not itself a benchmark dimension. 
Rather, it serves as a common interface that enables downstream evidence extraction. 
In the current implementation, defect evidence may be derived from count-based features, similarity patterns, dependency-graph structure, and context-state transitions, depending on defect type.

\subsection{Defect Taxonomy and Benchmark Instantiation}

ProcCtrlBench structures process evaluation around a reusable defect ontology. Let \(Z\) denote the set of defect types benchmarked by ProcCtrlBench, with each \(z_j \in Z\) corresponding to a specific execution defect. In this work, ProcCtrlBench covers eleven defect types organized into four top-level categories: Context Management, Tool-Use Efficiency, Workflow Architecture, and Tool-Ecosystem Consistency.
For each defect class, ProcCtrlBench extracts structured evidence
\[
x_j = \phi_j(T),
\]
 and then computes a continuous evidence score,
\[
s_j = f_j(x_j).
\]
We retain \(s_j\) as a continuous signal because many process defects are graded rather than naturally binary. When binary activation is needed, it is derived only as an intermediate interpretation layer:
\[
\text{finding}_j =
\begin{cases}
1, & s_j \ge \tau_j \land \neg \text{exempt}_j(T),\\
0, & \text{otherwise},
\end{cases}
\]
Exemption logic is important because some trajectories may superficially match a defect pattern while remaining semantically appropriate under task context, such as intended validation reruns or necessary long-lived instruction text.

A key design choice is that the ontology and the current instantiation should not be conflated. 
The ontology defines what kinds of failures ProcCtrlBench aims to evaluate; the detectors, thresholds, and scoring rules used here are one practical instantiation of that benchmark space.

\subsection{Calibrated Process Scoring}

Raw detector scores do not by themselves provide consistent semantics across defect types or task conditions. The same score magnitude may correspond to different empirical defect likelihoods under different execution contexts \citep{d2022underspecification}. 
ProcCtrlBench therefore maps benchmark evidence into calibrated posterior risk before producing scorecards.

For the \(j\)-th defect, let \(y_j \in \{0,1\}\) indicate whether the defect is truly present, let \(x_j\) denote the extracted evidence, and \(c\) denote task- and environment-level conditioning variables. The posterior defect risk is 
\[
p_j = P(y_j = 1 \mid x_j, c).
\]
For interpretability, posterior risks are mapped into semantic bands using thresholds \(\delta_w < \delta_e\):
\[
\text{sev}_j =
\begin{cases}
\text{error}, & p_j \ge \delta_e,\\
\text{warning}, & \delta_w \le p_j < \delta_e,\\
\text{none}, & p_j < \delta_w.
\end{cases}
\]

These labels are reporting aids rather than the benchmark score itself. ProcCtrlBench aggregates defect-level risks into dimension-level process-quality scores across the four benchmark dimensions, and then into an overall process-quality summary \(Q_{\text{def}}\).
Because the benchmark is intended primarily as a structured reporting tool, we emphasize the full scorecard over any single scalar.

\subsection{Control Preservation and Benchmark Output}

ProcCtrlBench uses control preservation to quantify execution-process quality during autonomous execution \citep{flournoy2020building}. . This matters because a trajectory may reach a correct outcome while
still being difficult to interpret, interrupt, repair, or safely halt.

We decompose control preservation into five subdimensions: interpretability, interruptibility, correctability, reversibility, and authority handoff. Together, these properties capture whether execution remains legible to an external observer, whether intervention is possible at meaningful points, whether local deviations can be repaired without restarting the trajectory, whether the system can return to a prior safe state, and whether decision authority can be returned when needed.

Let \(\psi(T)\) denote control-related features extracted from trajectory \(T\), and let \(F\) denote the set of benchmark findings. We define the aggregate control-preservation score as 
\[
CP = g(\psi(T), F).
\]
Control preservation is not merely the inverse of defect burden. A trajectory may display multiple process defects yet still preserve clear stage visibility, rollback structure, or meaningful intervention points. Conversely, a trajectory that appears superficially successful may preserve very little practical control. ProcCtrlBench therefore incorporates control preservation as an important component in the assessment of execution-process quality.

For compact comparison, ProcCtrlBench reports a summary score
\[
PB = \eta Q_{\text{def}} + (1 - \eta) CP, \quad 0 \le \eta \le 1.
\]
However, we stress that this scalar is secondary. The primary benchmark output is the full scorecard, which exposes defect-specific findings, dimension-level quality, control-preservation assessments, and coverage information.

Although ProcCtrlBench is proposed primarily as a benchmark, its calibrated process findings may also support execution-time governance. In this setting, benchmark findings and current control state can be used to regulate autonomy, trigger confirmation or rollback, or request handoff. We emphasize that this downstream use is not part of the benchmark definition itself; it is included only to illustrate that benchmark-calibrated process signals may also be operationally useful.

\section{Experiments}

We evaluate ProcCtrlBench through four questions: (1) whether the benchmark can be instantiated reliably from real execution trajectories; (2) whether calibrated scoring provides more stable benchmark semantics than direct thresholding; (3) whether ProcCtrlBench reveals system-level differences beyond outcome-based metrics; and (4) whether the benchmark remains informative across task scenarios, with exploratory analysis of auxiliary downstream utility.

\subsection{Experimental Setup}

We use a mixed evaluation set of 200 cases sampled from three coding-agent benchmarks: 100 cases from SWE-bench-Verified, 40 from AndroidBench, and 60 from TerminalBench.
This composition is intended to cover a range of coding scenarios, including repository-level bug fixing, terminal-centric tool use, and Android-oriented engineering tasks.
We evaluate 11 agent–model configurations spanning four agent families: Claude Code, Codex CLI, OpenCode, and Qoder.

For benchmark validation, we annotate the same 200-case subset at the process level.
Each trajectory is independently reviewed by two annotators with prior experience in AI coding or agent-based tool use, and disagreements are resolved by adjudication from a third reviewer when needed. 
Annotators assess defect presence under the four benchmark dimensions, record triggering evidence, and note exemption rationales where applicable. 
Annotations are process-oriented rather than outcome-oriented: repeated validation, persistent rule text, or long execution are labeled as defects only when they are semantically inappropriate under task context.

To separate detector tuning from evaluation, we split the 200 annotated cases into 80 development cases, 40 calibration cases, and 80 evaluation cases, stratified by benchmark source and task success.
Detector thresholds are initialized heuristically and refined on the development split only. 
Posterior risk mappings are fit on the calibration split and reported on the held-out evaluation split. 
Appendix~\ref{app:setup:agreement} reports inter-annotator agreement and adjudication statistics.

For each system, we retain complete execution logs rather than only final outputs. 
All logs are transformed by \(\Psi\) into standardized trajectories, after which ProcCtrlBench produces per-defect calibrated findings, dimension-level process-quality scores, control-preservation assessments, and overall benchmark scores.
Detector thresholds are initialized heuristically and refined on annotated trajectories, while exemption rules are applied after score computation and before posterior risk estimation. Posterior risks are estimated using simple Bayesian or frequency-smoothed updates for low-dimensional detectors and calibrated score-to-risk mappings for higher-dimensional detectors.

Detector performance is measured using precision, recall, F1, Average Precision, and AUROC. Calibration is measured using Expected Calibration Error (ECE). For system comparison, we report the overall ProcCtrlBench score \(PB\), dimension-level quality scores, control preservation \(CP\), and fragile-success rate.

\subsection{RQ1: Instantiation Reliability}

We first evaluate whether ProcCtrlBench defect classes can be instantiated with useful reliability from real execution trajectories. This is a prerequisite for meaningful benchmark use: if benchmarked defects cannot be grounded in trajectory evidence with reasonable consistency, downstream comparison becomes weakly supported.

Table~\ref{tab:instantiation_calibration} reports precision, recall, F1, AP, and AUROC for all 11 defect classes on the annotated set. Defects with strong local observability, such as Ghost Context, Duplicate Step, Dead Step, and Long Chain, achieve stronger performance. Structurally higher-level defects such as Wrapper Workflow, Context Coupling, and Weak Tool are more difficult to instantiate, which is expected because they depend more heavily on latent workflow structure and higher-level execution intent.

\begin{figure*}[t]
\centering
\begin{minipage}[t]{0.48\textwidth}
    \centering
    \includegraphics[width=\textwidth]{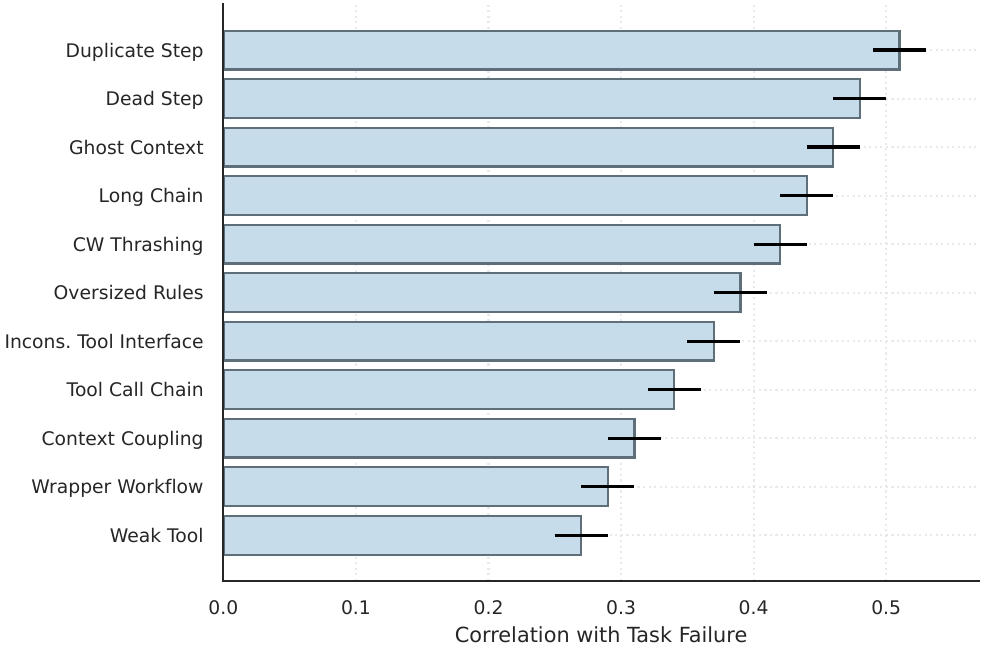}
    
    \small (a) Correlation between defect benchmarks and downstream task failure
\end{minipage}
\hfill
\begin{minipage}[t]{0.48\textwidth}
    \centering
    \includegraphics[width=\textwidth]{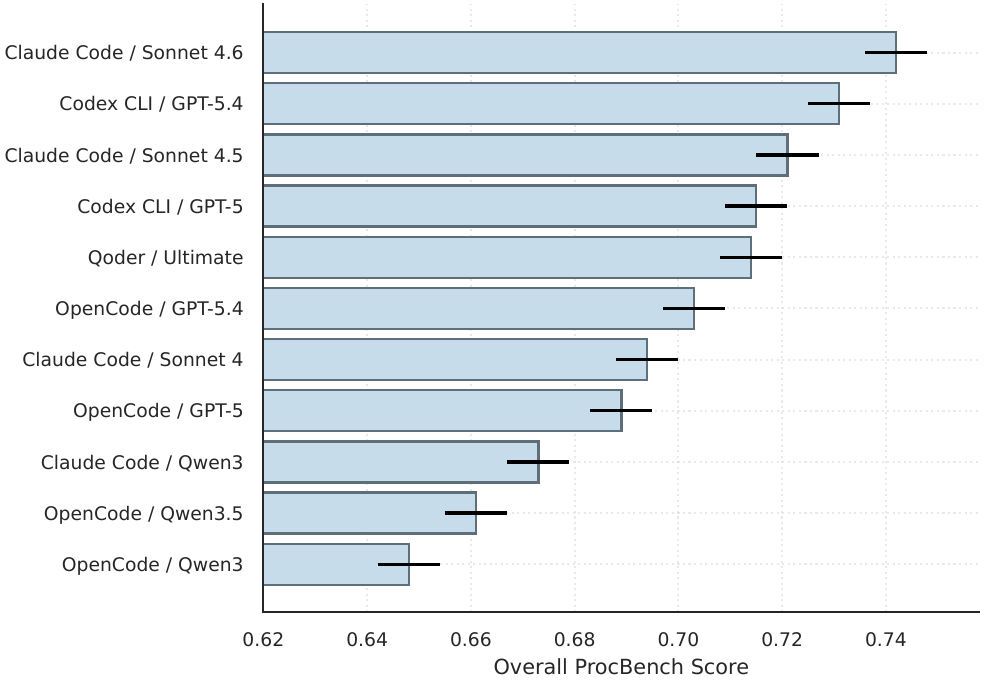}
    
    \small (b) Overall ProcCtrlBench score across agent--model configurations
\end{minipage}
\caption{Overview results for ProcCtrlBench. The left panel shows the association between defect benchmarks and downstream task failure. Long Chain and Dead Step show the strongest associations, while Context Window Thrashing, Context Coupling, Duplicate Step, and Ghost Context also exhibit meaningful positive correlations. The right panel shows the overall ProcCtrlBench score across the 11 evaluated agent--model configurations.}
\label{fig:overview_results}
\end{figure*}

We next examine whether ProcCtrlBench findings are behaviorally meaningful. Figure~\ref{fig:overview_results} (left) shows associations between each defect and downstream task failure. Long Chain and Dead Step show the strongest positive correlations, followed by Context Window Thrashing and Context Coupling. These patterns suggest that ineffective local progress, elongated execution, and degraded context handling are meaningfully related to execution breakdown.

Overall, the results suggest that several ProcCtrlBench defect classes can be instantiated with useful reliability, although observability varies by defect type and should be interpreted accordingly.

\begin{table*}[t]
\centering
\caption{Defect-level instantiation performance and calibration quality of ProcCtrlBench on the 200-case annotated set. The left block reports defect-level detection metrics, and the right block reports Expected Calibration Error (ECE) for the corresponding benchmark dimension. Higher is better for Precision, Recall, F1, AP, and AUROC; lower is better for ECE.}
\label{tab:instantiation_calibration}
\setlength{\tabcolsep}{4.0pt}
\renewcommand{\arraystretch}{1.08}
\begin{threeparttable}
\begin{adjustbox}{max width=\textwidth}
\begin{tabular}{l l c c c c c c | c c}
\toprule
\multicolumn{8}{c|}{\textbf{Defect-level Instantiation}} & \multicolumn{2}{c}{\textbf{Calibration (ECE)}} \\
\cmidrule(r){1-8} \cmidrule(l){9-10}
Group & Defect Benchmark & Trigger & Precision & Recall & F1 & AP & AUROC & Hard & Bayes \\
\midrule
\multirow{3}{*}{Context Management}
& Ghost Context   & 0.61 & 0.84 & 0.86 & 0.85 & 0.83 & 0.91 & \multirow{3}{*}{0.214} & \multirow{3}{*}{\textbf{0.118}} \\
& Oversized Rules & 0.43 & 0.81 & 0.79 & 0.80 & 0.79 & 0.88 & & \\
& CW Thrashing    & 0.47 & 0.78 & 0.80 & 0.79 & 0.77 & 0.86 & & \\
\midrule
\multirow{4}{*}{Tool-Use Efficiency}
& Duplicate Step  & 0.38 & 0.86 & 0.85 & 0.85 & 0.85 & 0.92 & \multirow{4}{*}{0.198} & \multirow{4}{*}{\textbf{0.103}} \\
& Tool Call Chain & 0.29 & 0.79 & 0.76 & 0.77 & 0.75 & 0.85 & & \\
& Dead Step       & 0.42 & 0.83 & 0.81 & 0.82 & 0.81 & 0.90 & & \\
& Long Chain      & 0.55 & 0.80 & 0.83 & 0.81 & 0.80 & 0.89 & & \\
\midrule
\multirow{2}{*}{Workflow Architecture}
& Wrapper Workflow & 0.21 & 0.61 & 0.57 & 0.59 & 0.56 & 0.71 & \multirow{2}{*}{0.271} & \multirow{2}{*}{\textbf{0.196}} \\
& Context Coupling & 0.26 & 0.64 & 0.59 & 0.61 & 0.58 & 0.73 & & \\
\midrule
\multirow{2}{*}{Tool Ecosystem}
& Incons. Tool Interface & 0.24 & 0.76 & 0.74 & 0.75 & 0.73 & 0.84 & \multirow{2}{*}{0.223} & \multirow{2}{*}{\textbf{0.134}} \\
& Weak Tool              & 0.17 & 0.55 & 0.52 & 0.53 & 0.50 & 0.68 & & \\
\midrule
Overall & - & - & - & - & - & - & - & 0.227 & \textbf{0.138} \\
\bottomrule
\end{tabular}
\end{adjustbox}
\begin{tablenotes}
\footnotesize
\item CW Thrashing: Context Window Thrashing. ECE: Expected Calibration Error.
\end{tablenotes}
\end{threeparttable}
\end{table*}

\subsection{RQ2: Calibration Quality}

We next evaluate whether calibrated scoring improves the empirical interpretability of ProcCtrlBench findings relative to direct thresholding. This question is important because ProcCtrlBench is intended to support cross-defect and cross-system comparison, where raw detector magnitudes may not carry comparable meaning.

Table~\ref{tab:instantiation_calibration} reports Expected Calibration Error for direct hard-threshold interpretation and hierarchical Bayesian calibration. Across all four defect dimensions, calibrated posterior risk yields lower ECE. The gains are largest in Workflow Architecture and Tool-Ecosystem Consistency, where evidence is more heterogeneous and observability is weaker. In these settings, a single hard threshold is especially brittle because the same detector magnitude may correspond to different empirical defect likelihoods under different conditions.

These results suggest that posterior-risk-based scoring provides more consistent empirical interpretation than raw trigger decisions in our annotated setting. 
Additional analyses indicate that warning- and error-level findings correspond to increasingly higher empirical defect frequency, supporting the interpretability of the severity bands; reliability diagrams in Appendix~\ref{fig:appendix_reliability} further show that calibrated risks more closely track empirical defect frequency than hard-threshold outputs. 
At the same time, these conclusions should be read in light of the current annotation scale and the lack of broader out-of-distribution validation.

\subsection{RQ3: Cross-System Benchmarking Beyond Outcome Metrics}

\begin{figure*}[t]
\centering
\includegraphics[width=0.97\textwidth]{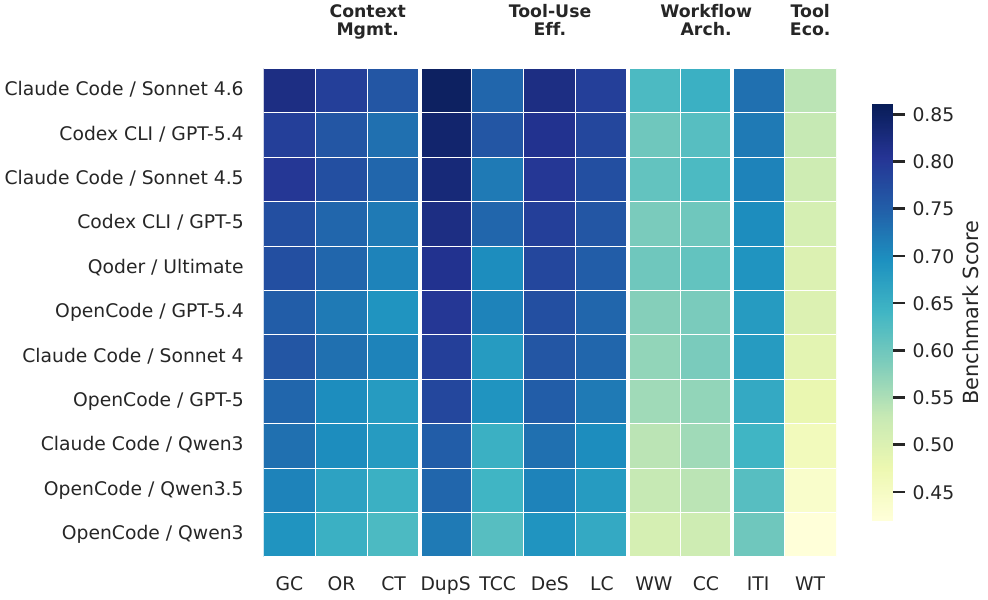}
\caption{Per-defect ProcCtrlBench profiles across the 11 evaluated systems. Rows denote systems and columns denote defect benchmarks. The heatmap highlights fine-grained differences beyond the overall score.}
\label{fig:defect-heatmap}
\end{figure*}

The main purpose of ProcCtrlBench is to compare coding-agent systems beyond final-outcome performance. We therefore evaluate all 11 agent–model configurations using the full ProcCtrlBench scorecard. Table~\ref{tab:main-scorecard} reports the overall ProcCtrlBench score, dimension-level quality scores, control preservation, and fragile-success rate. The right panel of Figure~\ref{fig:overview_results} summarizes the overall ranking, and Figure~\ref{fig:defect-heatmap} provides per-defect profiles. Several patterns emerge.

\begin{table*}[t]
\centering
\caption{Cross-agent and cross-model comparison under ProcCtrlBench on the 200-case mixed evaluation set. The left block reports overall scorecard metrics, and the right block reports the overall ProcCtrlBench score \(PB\) on each task source. Higher is better for \(PB\), \(Q_{\mathrm{ctx}}\), \(Q_{\mathrm{tool}}\), \(Q_{\mathrm{wf}}\), \(Q_{\mathrm{eco}}\), and \(CP\); lower is better for Fragile Success.}
\label{tab:main-scorecard}
\setlength{\tabcolsep}{4.2pt}
\renewcommand{\arraystretch}{1.10}
\begin{threeparttable}
\begin{adjustbox}{max width=\textwidth}
\begin{tabular}{c l c c c c c c c c c c}
\toprule
\multirow{2}{*}{Agent} & \multirow{2}{*}{Model} & \multicolumn{7}{c}{Overall Scorecard} & \multicolumn{3}{c}{Scenario-specific \(PB\)} \\
\cmidrule(lr){3-9} \cmidrule(lr){10-12}
& & \(PB\) & \(Q_{\mathrm{ctx}}\) & \(Q_{\mathrm{tool}}\) & \(Q_{\mathrm{wf}}\) & \(Q_{\mathrm{eco}}\) & \(CP\) & Fragile Succ. & Android & Terminal & SWE-bench \\
\midrule
\multirow{4}{*}{\makecell[c]{Claude\\Code}}
& Claude Sonnet 4.6 & \textbf{0.742} & \textbf{0.78} & 0.74 & \textbf{0.71} & 0.69 & \textbf{0.75} & \textbf{10.8\%} & 0.756 & 0.731 & 0.744 \\
& Claude Sonnet 4.5 & 0.721 & 0.76 & 0.71 & 0.69 & 0.67 & 0.73 & 12.1\% & 0.733 & 0.716 & 0.724 \\
& Claude Sonnet 4   & 0.694 & 0.72 & 0.69 & 0.66 & 0.65 & 0.70 & 15.4\% & 0.704 & 0.689 & 0.697 \\
& Qwen3 Coder Plus  & 0.673 & 0.70 & 0.66 & 0.64 & 0.63 & 0.68 & 17.3\% & 0.682 & 0.664 & 0.671 \\
\midrule
\multirow{2}{*}{\makecell[c]{Codex\\CLI}}
& gpt-5.4-0305-global & 0.731 & 0.74 & \textbf{0.75} & 0.68 & \textbf{0.70} & 0.72 & 11.7\% & 0.721 & \textbf{0.742} & 0.732 \\
& GPT-5 Global        & 0.715 & 0.72 & 0.74 & 0.67 & 0.68 & 0.71 & 12.9\% & 0.707 & 0.728 & 0.717 \\
\midrule
\multirow{4}{*}{\makecell[c]{Open\\Code}}
& gpt-5.4-0305-global & 0.703 & 0.71 & 0.72 & 0.65 & 0.67 & 0.69 & 14.6\% & 0.694 & 0.711 & 0.704 \\
& GPT-5 Global        & 0.689 & 0.69 & 0.71 & 0.64 & 0.66 & 0.68 & 15.8\% & 0.681 & 0.697 & 0.691 \\
& qwen3.5-plus        & 0.661 & 0.67 & 0.65 & 0.63 & 0.61 & 0.66 & 18.5\% & 0.654 & 0.672 & 0.664 \\
& Qwen3 Coder Plus    & 0.648 & 0.65 & 0.64 & 0.61 & 0.60 & 0.65 & 20.2\% & 0.641 & 0.659 & 0.651 \\
\midrule
Qoder & Ultimate mode & 0.714 & 0.73 & 0.70 & 0.69 & 0.66 & 0.71 & 13.6\% & 0.708 & 0.719 & 0.716 \\
\bottomrule
\end{tabular}
\end{adjustbox}
\begin{tablenotes}
\footnotesize
\item \(Q_{\mathrm{ctx}}\): Context Mgmt.; \(Q_{\mathrm{tool}}\): Tool-Use Eff.; \(Q_{\mathrm{wf}}\): Workflow Arch.; \(Q_{\mathrm{eco}}\): Tool Ecosystem.
\item Android: AndroidBench; Terminal: TerminalBench; SWE-bench: SWE-bench-Verified.
\end{tablenotes}
\end{threeparttable}
\end{table*}

First, ProcCtrlBench exposes meaningful differences in process profile across systems. Claude Code with Claude Sonnet 4.6 achieves the highest overall summary score, while Codex CLI is particularly competitive in tool-use efficiency. OpenCode variants remain competitive in some outcome-oriented settings but are weaker in workflow quality, tool-ecosystem consistency, control preservation, and fragile-success rate.

Second, process quality appears to depend jointly on model capability and agent orchestration. Within the same framework, stronger model variants generally improve the summary score and reduce fragile success. Across frameworks, similar model families behave differently once embedded in different agent infrastructures. This supports the view that process reliability is a property of the combined agent–model system rather than of the base model alone.

Third, ProcCtrlBench mainly adds diagnostic granularity beyond conventional outcome metrics. In most cases it does not radically overturn the overall ranking, but it helps distinguish systems with similar endpoint performance by revealing differences in process-level findings, workflow quality, and control preservation.
This is particularly visible in systems whose outcome scores remain competitive while their trajectories exhibit weaker process structure or higher fragile-success rates.

Taken together, these results suggest that ProcCtrlBench can support structured comparative evaluation beyond endpoint success alone, especially when users care about reliability, diagnosability, and governability in addition to task completion. 
Bootstrap analysis in Appendix~\ref{app:additional:bootstrap} further indicates that coarse-grained distinctions between stronger and weaker systems are stable under trajectory resampling, while several middle-ranked systems remain statistically close.

To contextualize these differences, we also compare ProcCtrlBench with a simpler outcome-only ranking baseline in Appendix~\ref{app:additional:baseline}. 
The comparison shows that ProcCtrlBench usually preserves the coarse ordering of stronger and weaker systems, but adds finer diagnostic separation in cases where endpoint performance is similar but execution quality differs. 
This reinforces the intended role of ProcCtrlBench as a complementary process-oriented evaluation layer rather than a wholesale replacement for outcome-based metrics.

\subsection{RQ4: Robustness Across Task Scenarios}

Scenario-specific results show that system strengths vary meaningfully by benchmark source. Claude Code with Sonnet 4.6 performs strongest on AndroidBench and SWE-bench-Verified, while Codex CLI with GPT-5.4 performs strongest on TerminalBench. Additional scenario-specific scores for all evaluated systems are reported in the appendix (Table~\ref{tab:scenario_pb}). This variation is expected because the three sources stress different execution properties: AndroidBench emphasizes engineering workflow structure, TerminalBench emphasizes command- and tool-centric interaction, and SWE-bench-Verified emphasizes repository-level debugging and repair. ProcCtrlBench is sensitive to these differences while preserving a broadly stable notion of overall process quality.

At the same time, ranking is not arbitrarily unstable across scenarios. Systems that perform strongly overall generally remain competitive across sources, although their margins differ. This suggests that ProcCtrlBench captures both scenario-sensitive behavior and a more global notion of process reliability.

Additional analyses in the appendix suggest that removing calibration weakens the semantic consistency of benchmark findings, while removing exemption logic increases false positives for defects such as Duplicate Step and Ghost Context. 
Exploratory analyses also suggest that ProcCtrlBench findings may have auxiliary downstream utility, including uncertainty-guided execution in high-risk trajectories, although this is not the primary focus of the present benchmark study.

\section{Limitations and Conclusion}
ProcCtrlBench has several limitations. First, its ontology is not exhaustive: the 11 defect classes introduced here are intended as a practical starting point for process-oriented benchmarking rather than a complete taxonomy of coding-agent failures. Second, some defect classes are only partially observable from execution traces, especially higher-level structural defects such as Wrapper Workflow, Context Coupling, and Weak Tool. Third, the current instantiation depends on process-level annotation, and although we now report annotation workflow, agreement, and data-splitting details, the scale of this annotation remains limited relative to the breadth of agent behaviors that may arise in practice. Fourth, calibrated risk estimation is based on a relatively small annotated set and may be sensitive to distribution shifts across task styles, tool ecosystems, and workflow regimes. Fifth, control preservation should be understood as an operational proxy for governability rather than a complete theory of human oversight. Finally, because the current study focuses on coding agents, additional validation is needed before extending strong conclusions to other agent domains.

In this paper, we introduced ProcCtrlBench, a benchmark for evaluating LLM coding agents beyond final outcomes. ProcCtrlBench organizes recurring execution defects into a reusable ontology, instantiates them from standardized execution trajectories, and reports calibrated scorecards that incorporate control preservation to quantify execution-process quality. Experiments on 200 annotated cases sampled from AndroidBench, TerminalBench, and SWE-bench-Verified suggest that several defect types can be instantiated with useful reliability, that calibration improves empirical interpretability relative to direct thresholding, and that process-aware scorecards provide diagnostic insight beyond conventional outcome-based evaluation. More broadly, these results support the view that evaluating coding agents should account not only for whether a task succeeds, but also for how execution unfolds.

\bibliography{iclr2026_conference}
\bibliographystyle{iclr2026_conference}

\appendix

\section{Detailed Benchmark Definitions}
\label{app:definitions}

This appendix provides expanded definitions and boundary conditions for the process defects introduced in the main text. ProcCtrlBench is designed to benchmark recurring execution pathologies that plausibly arise in realistic coding-agent deployment, rather than assigning arbitrary labels to trajectories after the fact. In long-horizon coding tasks, many failures emerge not from a single catastrophic mistake, but from the interaction between context pressure, tool interfaces, workflow composition, and execution control. The benchmark definitions below therefore emphasize three aspects: what the defect is, where its practical boundary lies, and why it matters for process-level evaluation.

\subsection{Context Management}
\label{app:definitions:context}

This dimension concerns the preservation, saturation, and degradation of contextual resources during execution. In coding-agent settings, context is a constrained operational resource that must support task continuity, tool coordination, and error recovery over multiple steps.

\paragraph{Ghost Context.}
Ghost Context refers to redundant, outdated, or already summarized content that remains in context for an extended period and continues to consume the token budget. Let \(u\) denote a context segment, with occupancy ratio \(\rho(u)\), downstream reference rate \(\kappa(u)\), and persistence duration \(\tau(u)\). This defect typically appears as high occupancy, low reference, and long persistence. Necessary long-lived rule text, explicitly retained summaries, and persistent memory are excluded.

In practice, Ghost Context often arises when agents retain raw file excerpts, command outputs, or earlier observations after their informational role has already been superseded by a concise summary or a more recent state. The risk is not merely wasted tokens, but also increased anchoring on stale evidence and reduced room for more relevant context.

\paragraph{Oversized Rules.}
Oversized Rules refers to excessively large system prompts, policy text, or instruction blocks that impose a persistent context burden at every step. Unlike Ghost Context, this defect concerns static rule overhead rather than dynamically retained redundant content.

This defect often appears in deployments where safety rules, style guidelines, workflow instructions, and tool documentation are accumulated into a single large prompt prefix. Even when each component is individually reasonable, the combined overhead can consume a substantial fraction of the context window and reduce effective space for task-specific reasoning.

\paragraph{Context Window Thrashing.}
Context Window Thrashing denotes a high-frequency pattern of heavy information loading followed by rapid saturation and short-term compression, in which information is compressed or discarded before it can form a stable reasoning chain.

The core issue is not large context alone, but unstable context turnover. Agents repeatedly acquire, compress, and replace information without preserving enough connective structure for durable reasoning. This often leads to rediscovery behavior, weak hypothesis continuity, and repeated local reconstruction of facts that had already been observed earlier in the trajectory.

\subsection{Tool-Use Efficiency}
\label{app:definitions:tool}

This dimension concerns repeated calls, local looping, ineffective steps, and disproportionately long execution chains in tool use. In deployed coding agents, tool interaction is the primary interface between model reasoning and environment action, so failures in this dimension are often directly visible as activity without meaningful progress.

\paragraph{Duplicate Step.}
Duplicate Step refers to the repeated execution of highly similar tool calls despite the absence of substantive state change. Validation reruns, time-varying commands, and intended batch processing are exempt.

This defect commonly arises when the agent lacks sufficient evidence to decide whether a previous call resolved its uncertainty, and therefore repeats low-information actions such as reopening the same file span, rerunning near-identical search commands, or issuing repeated status checks without meaningful environmental change. The main signal is weak local progress rather than repetition alone.

\paragraph{Tool Call Chain.}
Tool Call Chain captures cyclic or oscillatory patterns in local call sequences, such as repeated retries of the same tool or loops of the form \(A \rightarrow B \rightarrow A \rightarrow B\).

Its practical significance lies in local entrapment: the agent remains active, but its tool sequence no longer reflects strategic escalation or replanning. Compared with Duplicate Step, Tool Call Chain emphasizes the structure of repeated local interaction rather than similarity of individual calls alone.

\paragraph{Dead Step.}
Dead Step describes a step whose action is executed but whose result neither enters subsequent messages nor affects later tool inputs, file modifications, branch decisions, or state updates.

Dead Steps indicate weak causal integration within the trajectory. Tool outputs may be retrieved, inspected, or generated, but if they exert no downstream influence, the step becomes operationally inert. This matters because it reflects not only inefficiency, but also poor coupling between observation and subsequent action.

\paragraph{Long Chain.}
Long Chain concerns whether the overall execution path is disproportionately long. Its core signal is not the step count itself, but persistent elongation in the absence of effective task decomposition, stage control, and checkpointing.

Long trajectories are not inherently defective, especially in repository-level debugging or multi-stage validation tasks. ProcCtrlBench therefore treats Long Chain as a structured burden signal rather than a length threshold alone. It is activated when execution keeps expanding without corresponding consolidation of progress.

\subsection{Workflow Architecture}
\label{app:definitions:workflow}

This dimension evaluates whether workflow organization provides genuine structural benefit. Modern coding agents increasingly employ skills, wrappers, staged pipelines, and sub-agent decomposition, but such structure is only beneficial when it improves validation, routing, fault tolerance, or control.

\paragraph{Wrapper Workflow.}
Wrapper Workflow refers to a workflow or skill that consistently performs only one-step pass-through behavior without adding value through validation, aggregation, routing, or fault tolerance.

The defect does not target modularity itself. Rather, it targets architectural indirection that increases complexity while contributing little operational benefit. Thin wrappers may still be acceptable when they enforce policy or standardize interfaces, but repeated pass-through behavior without clear functional gain is benchmarked as structural overhead.

\paragraph{Context Coupling.}
Context Coupling refers to excessively deep context sharing and unclear boundaries across agents or workflow units, often manifested as bidirectional dependencies, ping-pong alternation, or strongly connected components.

Some degree of context sharing is expected in coordinated systems. The defect applies when sharing becomes so deep that unit boundaries lose meaning, local reasoning becomes difficult to isolate, and errors propagate across workflow components. This reduces diagnosability and weakens the ability to correct local deviations without broader trajectory disruption.

\subsection{Tool-Ecosystem Consistency}
\label{app:definitions:ecosystem}

This dimension concerns inconsistency within the surrounding tool ecosystem. Execution quality depends not only on model capability and workflow logic, but also on whether tools are discoverable, distinguishable, and interface-consistent for model-driven use.

\paragraph{Inconsistent Tool Interface.}
Inconsistent Tool Interface describes substantial inconsistency among functionally similar tools in parameter naming, parameter types, output structure, or error-return formats.

This defect is especially common in organically grown tool ecosystems. For LLM agents, such inconsistencies are more than minor usability issues: they make invocation patterns harder to generalize, increase malformed calls, and encourage fallback to less suitable but more familiar alternatives.

\paragraph{Weak Tool.}
Weak Tool refers to a tool that exists but is difficult for the model to discover and use reliably, for example when its invocation rate remains persistently low in clearly appropriate settings and it is often replaced by a more intuitive but only partially overlapping alternative.

The main issue is not capability absence, but ecosystem legibility. A tool may be technically available yet underused because its name, description, or interface fails to make its function salient to the model. In such cases, the benchmark captures weakness of exposure and practical accessibility rather than intrinsic tool quality alone.

\section{Experimental Setup Details}
\label{app:setup}

\subsection{Annotation Protocol}
\label{app:setup:annotation}

To evaluate benchmark instantiation quality, we annotate a subset of 200 cases sampled from the full corpus using reviewers with prior experience in AI coding or agent-based tool use. 
Annotators assess the presence of defects under the four benchmark dimensions---Context Management, Tool-Use Efficiency, Workflow Architecture, and Tool-Ecosystem Consistency---and record triggering evidence together with exemption rationales where needed. Disagreements are resolved through dual review and adjudication.

The annotation protocol is process-oriented rather than outcome-oriented. Annotators are instructed to judge whether a suspicious pattern constitutes a genuine process defect under the task context, rather than labeling surface repetition or failure alone. For example, repeated validation is not labeled as Duplicate Step when intervening edits make reruns appropriate, and persistent rule text is not labeled as Ghost Context when it serves an ongoing coordination or control role. This design is important because many process defects are only weakly reflected in final task success or failure.

Each annotated trajectory is reviewed as a causal execution process rather than as a collection of isolated events. In addition to binary presence judgments, annotators record brief evidence notes and exemption rationales where applicable. These notes support adjudication and help preserve the local semantic context needed for process-level evaluation.

\begin{table}[t]
\centering
\caption{Scenario-specific ProcCtrlBench scores across benchmark sources. Systems are sorted by overall mixed-set ProcCtrlBench score \(PB\).}
\label{tab:scenario_pb}
\setlength{\tabcolsep}{4.2pt}
\renewcommand{\arraystretch}{1.08}
\begin{threeparttable}
\begin{adjustbox}{width=\columnwidth}
\begin{tabular}{l c c c c}
\toprule
System & Overall & Android & Terminal & SWE-bench \\
\midrule
Claude Code / Sonnet 4.6           & 0.742 & 0.756 & 0.731 & 0.744 \\
Codex CLI / GPT-5.4                & 0.731 & 0.721 & 0.742 & 0.732 \\
Claude Code / Sonnet 4.5           & 0.721 & 0.733 & 0.716 & 0.724 \\
Codex CLI / GPT-5                  & 0.715 & 0.707 & 0.728 & 0.717 \\
Qoder / Ultimate                   & 0.714 & 0.708 & 0.719 & 0.716 \\
OpenCode / GPT-5.4                 & 0.703 & 0.694 & 0.711 & 0.704 \\
Claude Code / Sonnet 4             & 0.694 & 0.704 & 0.689 & 0.697 \\
OpenCode / GPT-5                   & 0.689 & 0.681 & 0.697 & 0.691 \\
Claude Code / Qwen3 Coder Plus     & 0.673 & 0.682 & 0.664 & 0.671 \\
OpenCode / Qwen3.5                 & 0.661 & 0.654 & 0.672 & 0.664 \\
OpenCode / Qwen3 Coder Plus        & 0.648 & 0.641 & 0.659 & 0.651 \\
\bottomrule
\end{tabular}
\end{adjustbox}
\end{threeparttable}
\end{table}

\subsection{Dataset and Annotation Statistics}
\label{app:setup:data}

The full task pool contains AI coding and multi-step tool-use trajectories with substantial variation in execution horizon, context length, number of tools, and workflow complexity. For detector validation and calibration, we construct an annotated subset of 200 cases sampled from AndroidBench, TerminalBench, and SWE-bench-Verified. This subset is sampled to cover short-, medium-, and long-horizon tasks, as well as both tool-light and tool-rich environments. We further ensure diversity across the four benchmark dimensions so that context-, tool-, workflow-, and ecosystem-related defects are all represented.

This diversity is important because process defects do not arise uniformly across task types. Compact terminal tasks more clearly expose local redundancy and inefficient tool use, while longer software-engineering tasks more often surface context accumulation, structural drift, and workflow coupling. Sampling across AndroidBench, TerminalBench, and SWE-bench-Verified therefore reduces the risk that ProcCtrlBench overfits to a single execution regime.

Each annotated trajectory is reviewed at the process level rather than solely at the endpoint. For each defect type, annotators determine whether the targeted defect is present, absent, or exempt under the task context.
This allows detector outputs to be evaluated against process-level human judgment rather than final outcome alone.

\subsection{Hyperparameter and Implementation Notes}
\label{app:setup:implementation}

Detector thresholds are initialized using heuristic rules and pilot inspection on a development subset, and are then refined against the annotated trajectories to balance precision and recall. Exemption rules are applied after score computation and before posterior risk estimation.

This design reflects the fact that process defects rarely admit a single universal operating point. A threshold suitable for short terminal trajectories may be too aggressive for repository-scale debugging, where longer chains or repeated validation steps can be semantically appropriate. Threshold refinement therefore aligns initial structural assumptions with the empirical semantics of real trajectories.

For warning/error grading, the warning threshold \(\delta_w\) and error threshold \(\delta_e\) are selected to preserve monotonic separation among low-, medium-, and high-risk findings on the annotated set. Posterior risk estimation uses different practical implementations depending on detector family. Low-dimensional count-based detectors use simple Bayesian updates or frequency-smoothed empirical estimation, while higher-dimensional detectors use calibrated score-to-risk mappings on held-out annotation data. Shared priors are specified at the defect-family level when beneficial, with class-specific adjustment retained for defect classes that exhibit sufficiently distinct evidence distributions.

Exemption rules are especially important because real coding-agent trajectories contain many superficially suspicious but contextually appropriate patterns. Applying exemptions after score computation but before posterior risk estimation allows ProcCtrlBench to preserve raw evidence while reducing false positives for semantically valid behavior.

\begin{table*}[t]
\centering
\caption{Conventional outcome metrics and ProcCtrlBench scorecard side by side across the 11 evaluated systems. Rank Shift is computed as Outcome Rank minus ProcCtrlBench Rank; positive values indicate that a system ranks better under ProcCtrlBench than under conventional outcome metrics.}
\label{tab:appendix_outcome_vs_pb}
\setlength{\tabcolsep}{4.2pt}
\renewcommand{\arraystretch}{1.08}
\begin{threeparttable}
\begin{adjustbox}{max width=\textwidth}
\begin{tabular}{l c c c c c c}
\toprule
System & Outcome Score & Outcome Rank & \(PB\) & \(PB\) Rank & Rank Shift & Fragile Success \\
\midrule
Claude Code / Sonnet 4.6        & 0.768 & 1 & 0.742 & 1 & 0  & 10.8\% \\
Codex CLI / GPT-5.4             & 0.759 & 2 & 0.731 & 2 & 0  & 11.7\% \\
Claude Code / Sonnet 4.5        & 0.752 & 3 & 0.721 & 3 & 0  & 12.1\% \\
Codex CLI / GPT-5               & 0.748 & 4 & 0.715 & 4 & 0  & 12.9\% \\
Qoder / Ultimate                & 0.744 & 5 & 0.714 & 5 & 0  & 13.6\% \\
OpenCode / GPT-5.4              & 0.745 & 4 & 0.703 & 6 & -2 & 14.6\% \\
Claude Code / Sonnet 4          & 0.733 & 7 & 0.694 & 7 & 0  & 15.4\% \\
OpenCode / GPT-5                & 0.731 & 8 & 0.689 & 8 & 0  & 15.8\% \\
Claude Code / Qwen3 Coder Plus  & 0.724 & 9 & 0.673 & 9 & 0  & 17.3\% \\
OpenCode / Qwen3.5              & 0.719 & 10 & 0.661 & 10 & 0 & 18.5\% \\
OpenCode / Qwen3 Coder Plus     & 0.712 & 11 & 0.648 & 11 & 0 & 20.2\% \\
\bottomrule
\end{tabular}
\end{adjustbox}
\begin{tablenotes}
\footnotesize
\item Outcome Score can be instantiated as task-success rate, test-pass rate, or a normalized outcome 

composite depending on logging availability.
\end{tablenotes}
\end{threeparttable}
\end{table*}

\begin{table*}[t]
\centering
\caption{Annotation agreement before adjudication on the 200-case annotated subset. Agreement is reported at the defect level. Higher is better.}
\label{tab:annotation_agreement}
\setlength{\tabcolsep}{4.5pt}
\renewcommand{\arraystretch}{1.08}
\begin{threeparttable}
\begin{adjustbox}{max width=\textwidth}
\begin{tabular}{l c c c c}
\toprule
Defect & Raw Agreement & Cohen's $\kappa$ & Exemption Rate & Adjudication Rate \\
\midrule
Ghost Context              & 0.87 & 0.73 & 0.06 & 0.12 \\
Oversized Rules            & 0.84 & 0.68 & 0.04 & 0.15 \\
Context Window Thrashing   & 0.82 & 0.64 & 0.05 & 0.17 \\
Duplicate Step             & 0.89 & 0.77 & 0.08 & 0.10 \\
Tool Call Chain            & 0.83 & 0.66 & 0.07 & 0.16 \\
Dead Step                  & 0.86 & 0.72 & 0.05 & 0.13 \\
Long Chain                 & 0.88 & 0.75 & 0.03 & 0.11 \\
Wrapper Workflow           & 0.76 & 0.52 & 0.05 & 0.22 \\
Context Coupling           & 0.78 & 0.55 & 0.04 & 0.21 \\
Inconsistent Tool Interface& 0.81 & 0.61 & 0.03 & 0.18 \\
Weak Tool                  & 0.74 & 0.47 & 0.06 & 0.24 \\
\midrule
Overall                    & 0.83 & 0.65 & 0.05 & 0.17 \\
\bottomrule
\end{tabular}
\end{adjustbox}
\end{threeparttable}
\end{table*}

\subsection{Annotation Agreement and Data Splits}
\label{app:setup:agreement}

To improve transparency of the annotation process, we report the annotation workflow and data partition used in the current instantiation of ProcCtrlBench. 
Each trajectory in the 200-case annotated subset is independently labeled by two annotators. 
When the two annotators disagree on the presence, absence, or exemption status of a defect, the case is sent to a third reviewer for adjudication. 
Agreement is computed before adjudication at the defect level.

The annotated set is partitioned into 80 development cases, 40 calibration cases, and 80 held-out evaluation cases. 
The split is stratified by benchmark source (AndroidBench, TerminalBench, SWE-bench-Verified) and by endpoint task success/failure, so that calibration and final evaluation are not dominated by a single task regime.

We emphasize that agreement varies by defect type. 
Locally observable defects such as Duplicate Step and Long Chain show higher consistency, while structurally higher-level defects such as Wrapper Workflow and Weak Tool exhibit lower agreement. 
This pattern is consistent with the detector-performance trends reported in Table~\ref{tab:instantiation_calibration}.

\section{Additional Results}

\subsection{Scenario-Specific Scores}

Table~\ref{tab:scenario_pb} reports the scenario-specific ProcCtrlBench scores for all evaluated systems on AndroidBench, TerminalBench, and SWE-bench-Verified. The results show that process quality is not monolithic across task regimes. Claude Code with Sonnet 4.6 remains strongest overall and performs best on AndroidBench and SWE-bench-Verified, while Codex CLI with GPT-5 achieves the highest score on TerminalBench. This pattern suggests that different agent--model systems exhibit different strengths under different execution environments.

At the same time, the overall ordering is not completely unstable across sources. Systems that rank strongly under the mixed evaluation set generally remain competitive across individual benchmarks, although the margin of advantage varies by scenario. This indicates that ProcCtrlBench captures both global process quality and scenario-specific sensitivity, rather than collapsing to either benchmark-specific noise or an over-smoothed average.

\subsection{Calibration Semantics}

Additional analyses indicate that the warning and error bands carry meaningful empirical semantics: error-level findings correspond to the highest empirical defect frequency, warning-level findings occupy an intermediate regime, and non-triggered findings correspond to the lowest. This supports the use of semantic severity labels as an interpretable reporting layer over calibrated risk.

\subsection{Exemption Rules and Auxiliary Utility}

Additional analyses suggest that exemption rules are important for reducing false positives, particularly for Duplicate Step and Ghost Context. We also observe that ProcCtrlBench findings may have auxiliary downstream utility, including uncertainty-guided execution in high-risk trajectories.

\subsection{Outcome-Based Ranking vs. ProcCtrlBench Ranking}

Table~\ref{tab:annotation_agreement} shows that agreement is highest for locally observable defects and lowest for structurally semantic defects, mirroring the relative instantiation difficulty observed in Table~\ref{tab:instantiation_calibration}.

\begin{figure*}[t]
\centering
\includegraphics[width=0.90\textwidth]{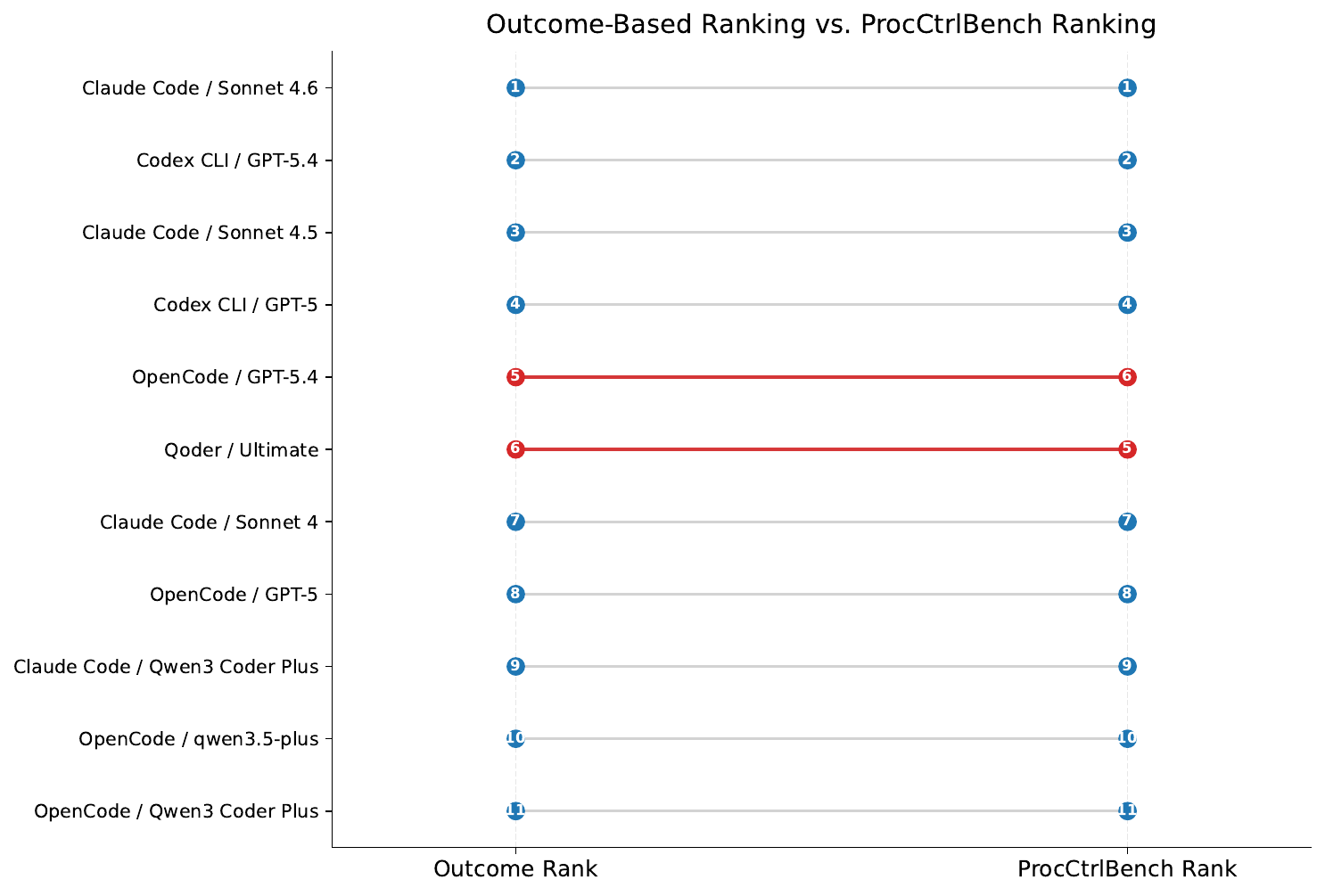}
\caption{Comparison between conventional outcome-based ranking and ProcCtrlBench ranking across the 11 evaluated systems. Lines indicate rank shifts from outcome-based evaluation to ProcCtrlBench.}
\label{fig:appendix_rank_compare}
\end{figure*}

To clarify how ProcCtrlBench differs from conventional evaluation, we compare system ranking under outcome-based metrics and under ProcCtrlBench. Figure~\ref{fig:appendix_rank_compare} visualizes rank shifts across the 11 evaluated systems, and Table~\ref{tab:appendix_outcome_vs_pb} reports the corresponding scores and fragile-success rates.

The comparison shows that ProcCtrlBench does not merely rescale outcome performance. Some systems with similar final success differ materially in process quality and control preservation, while some systems with competitive outcome scores still exhibit elevated fragile-success rates. These results support the claim that ProcCtrlBench captures execution properties that are weakly reflected in conventional outcome-only benchmarks.

\subsection{Comparison to a Simpler Outcome-Only Baseline}
\label{app:additional:baseline}

A natural baseline is to rank systems using only endpoint outcome metrics, such as task success or test-pass rate. 
Table~\ref{tab:appendix_outcome_vs_pb} compares this baseline with ProcCtrlBench. 
The comparison shows that ProcCtrlBench does not radically reorder the entire leaderboard; instead, it mainly adds diagnostic granularity by separating systems with similar outcome scores but different process profiles. 
This behavior is desirable for a process-oriented benchmark: the goal is not to replace endpoint evaluation, but to identify execution differences that endpoint metrics alone do not expose.

We summarize this effect in two ways. 
First, ProcCtrlBench preserves the top-3 systems under outcome-only ranking, indicating that stronger endpoint performance remains broadly aligned with stronger process quality at the coarse level. 
Second, ProcCtrlBench penalizes systems with elevated fragile-success rates and weaker workflow/control characteristics, even when their outcome scores remain competitive. 
OpenCode / GPT-5.4 is the clearest example in our current setting.

\subsection{Bootstrap Stability Analysis}
\label{app:additional:bootstrap}

\begin{figure*}[t]
\centering
\includegraphics[width=0.95\textwidth]{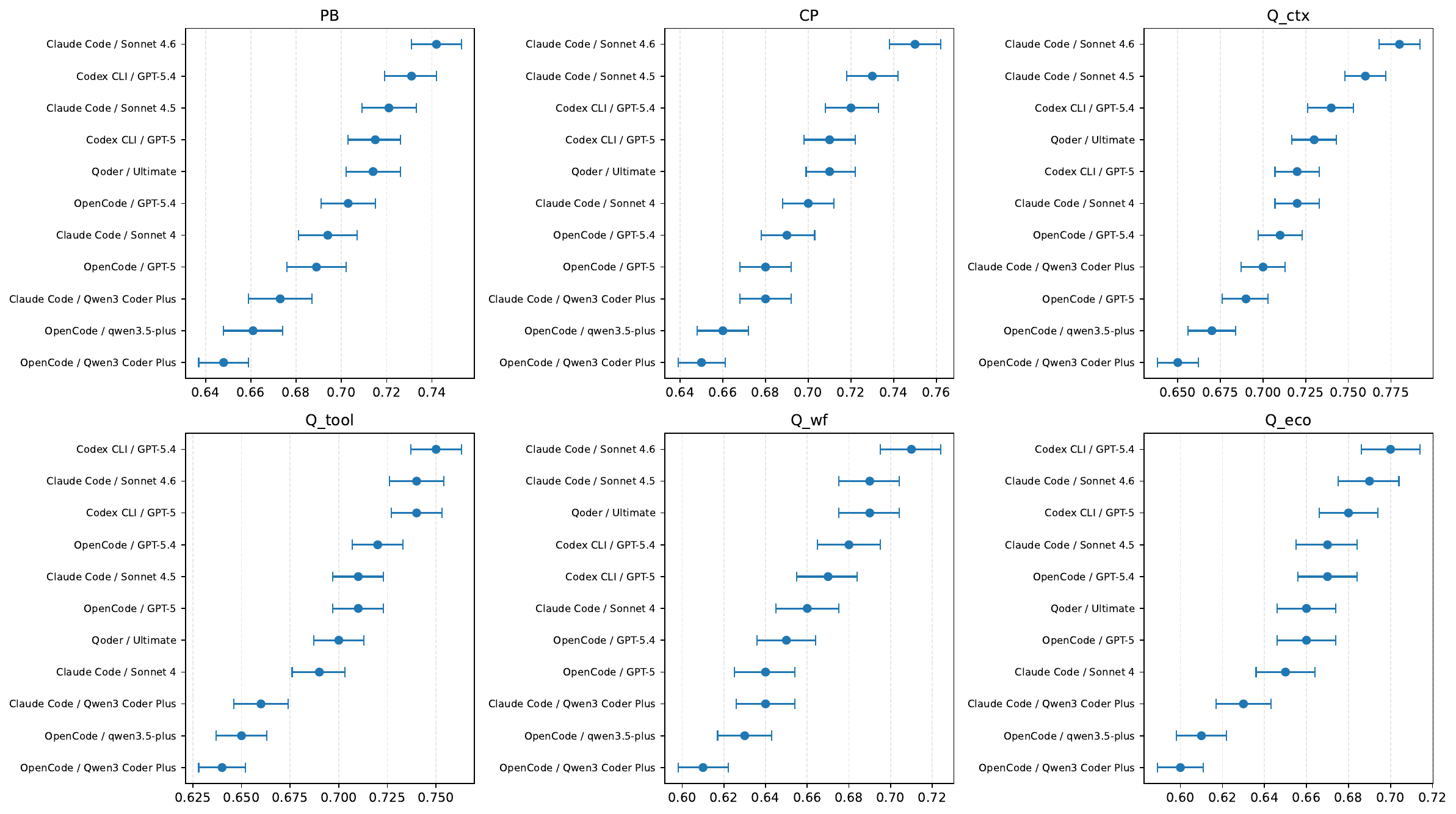}
\caption{Bootstrap confidence intervals for overall ProcCtrlBench score, control preservation, and dimension-level process-quality scores across the 11 evaluated systems. Error bars indicate 95\% bootstrap confidence intervals.}
\label{fig:appendix_bootstrap_ci}
\end{figure*}

We assess the stability of ProcCtrlBench under trajectory resampling using stratified bootstrap over the 200-case mixed evaluation set. Figure~\ref{fig:appendix_bootstrap_ci} reports bootstrap confidence intervals for overall ProcCtrlBench score, control preservation, and dimension-level process-quality scores. Table~\ref{tab:appendix_rank_stability} summarizes mean rank, rank standard deviation, Top-1 frequency, and Top-3 frequency across bootstrap replicates.

The results suggest that coarse-grained distinctions between stronger and weaker systems are stable, while several middle-ranked systems remain statistically close. This indicates that ProcCtrlBench is informative for structured comparative evaluation, but small differences between near-tied systems should be interpreted with caution.

\begin{table*}[t]
\centering
\caption{Bootstrap ranking stability of ProcCtrlBench across the 11 evaluated systems on the mixed evaluation set. Values are computed over stratified bootstrap replicates.}
\label{tab:appendix_rank_stability}
\setlength{\tabcolsep}{4.8pt}
\renewcommand{\arraystretch}{1.08}
\begin{threeparttable}
\begin{adjustbox}{max width=\textwidth}
\begin{tabular}{l c c c c}
\toprule
System & Mean Rank & Rank Std & Top-1 Freq. & Top-3 Freq. \\
\midrule
Claude Code / Sonnet 4.6        & 1.42 & 0.68 & 0.63 & 0.98 \\
Codex CLI / GPT-5.4             & 2.11 & 0.81 & 0.22 & 0.94 \\
Claude Code / Sonnet 4.5        & 2.87 & 0.93 & 0.10 & 0.81 \\
Codex CLI / GPT-5               & 4.26 & 1.17 & 0.03 & 0.31 \\
Qoder / Ultimate                & 4.39 & 1.12 & 0.02 & 0.28 \\
OpenCode / GPT-5.4              & 5.92 & 1.35 & 0.00 & 0.07 \\
Claude Code / Sonnet 4          & 6.41 & 1.28 & 0.00 & 0.03 \\
OpenCode / GPT-5                & 7.18 & 1.41 & 0.00 & 0.01 \\
Claude Code / Qwen3 Coder Plus  & 8.84 & 1.22 & 0.00 & 0.00 \\
OpenCode / Qwen3.5              & 9.74 & 0.88 & 0.00 & 0.00 \\
OpenCode / Qwen3 Coder Plus     & 10.86 & 0.42 & 0.00 & 0.00 \\
\bottomrule
\end{tabular}
\end{adjustbox}
\end{threeparttable}
\end{table*}

\subsection{Inter-Defect Correlation Structure}

\begin{figure*}[t]
\centering
\includegraphics[width=0.82\textwidth]{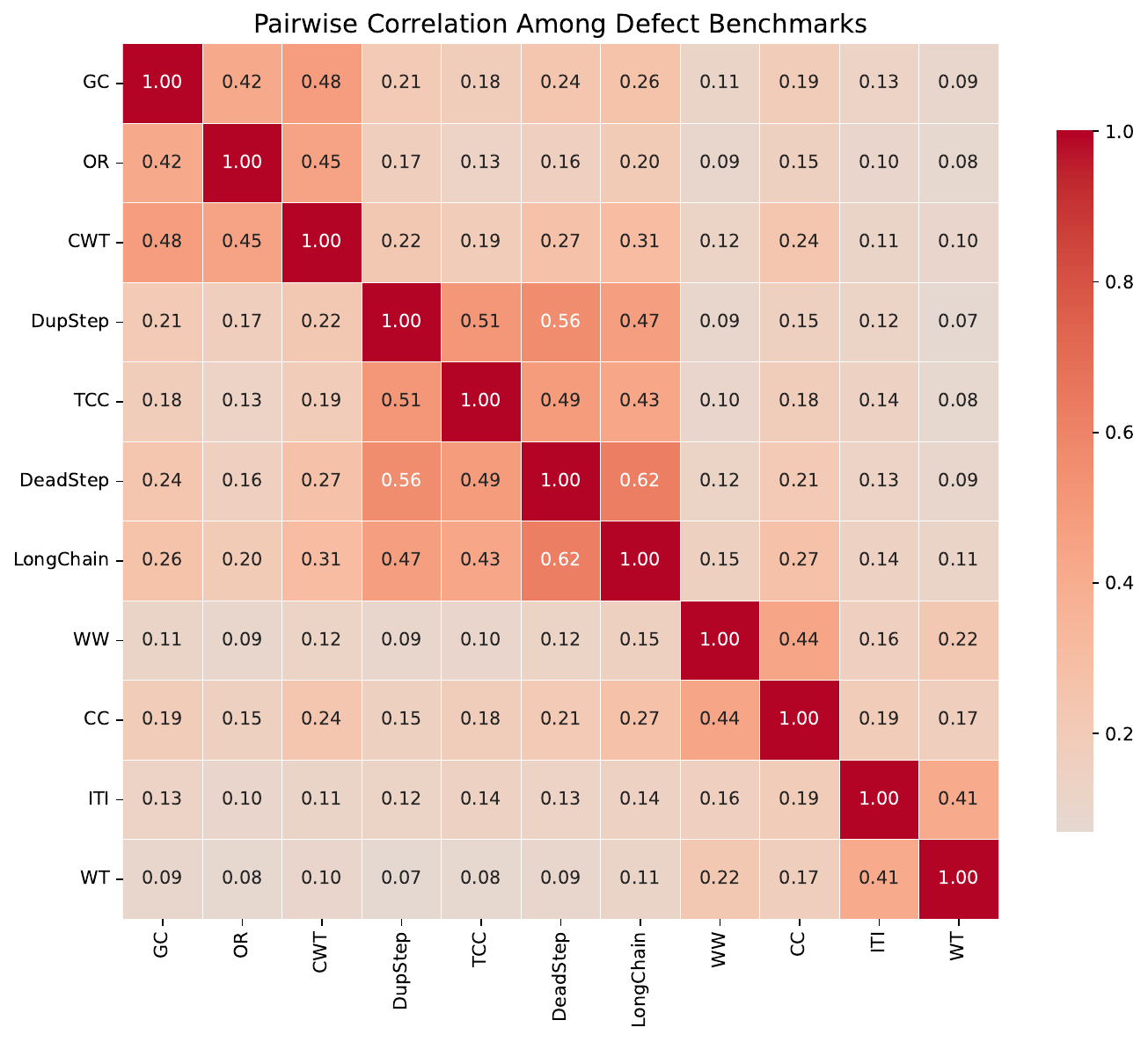}
\caption{Pairwise correlation matrix among defect-level posterior risks on the mixed evaluation set. Moderate correlations appear within some efficiency-related defects, while many cross-dimension pairs remain weaker.}
\label{fig:appendix_defect_corr}
\end{figure*}

To examine whether ProcCtrlBench defects are excessively redundant, we compute pairwise correlations among defect-level posterior risks across the mixed evaluation set. Figure~\ref{fig:appendix_defect_corr} shows the resulting correlation matrix.

Several efficiency-related defects, especially Long Chain and Dead Step, exhibit moderate positive association, which is expected because elongated execution often co-occurs with locally ineffective steps. However, correlations across dimensions are generally weaker, suggesting that ProcCtrlBench does not collapse into a single latent defect signal. This supports the scorecard design in which multiple partially related process defects are reported together rather than summarized as a raw defect count.

\subsection{Calibration Reliability Diagrams}

\begin{figure*}[t]
\centering
\includegraphics[width=0.95\textwidth]{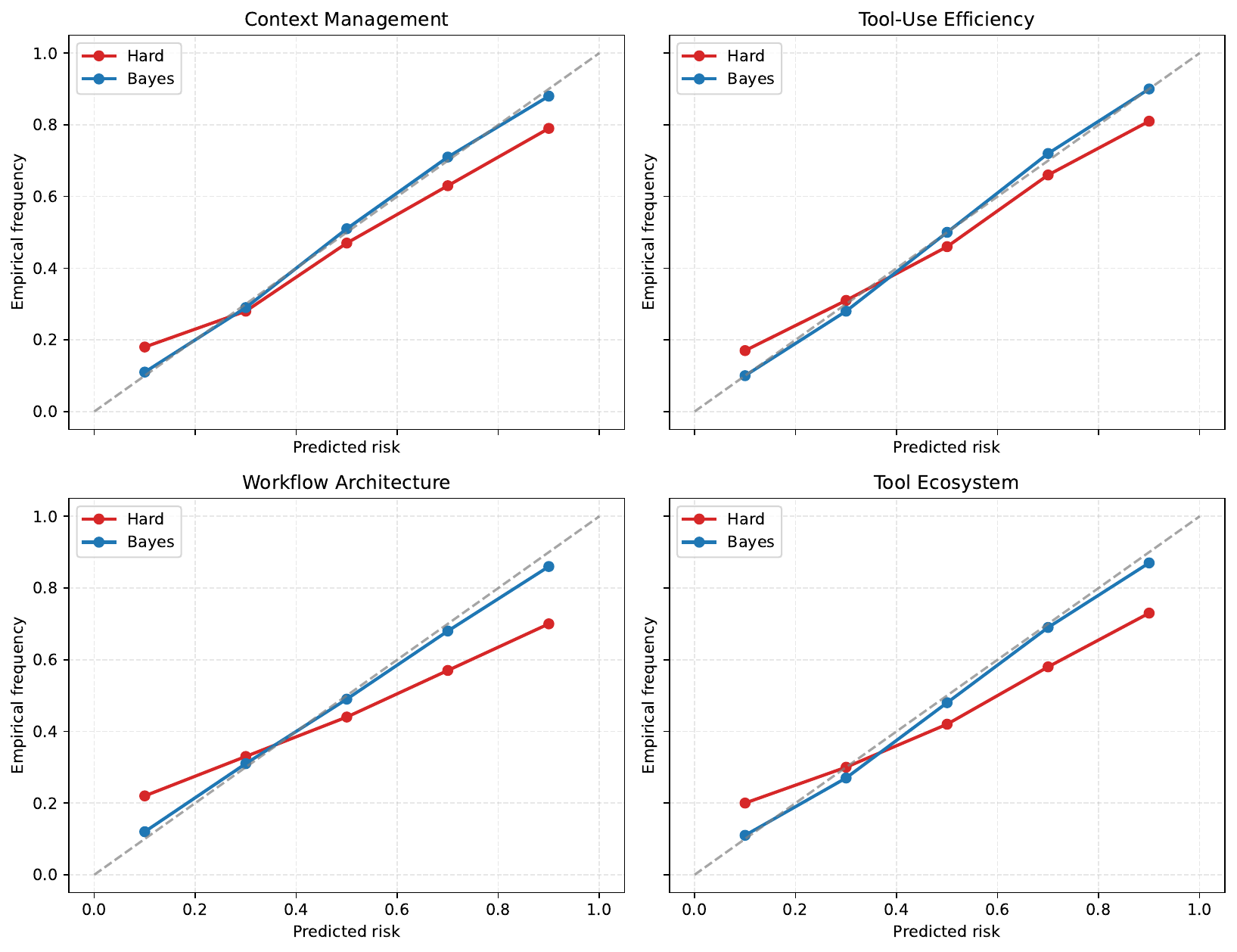}
\caption{Reliability diagrams for hard-threshold interpretation and calibrated posterior risk across the four benchmark dimensions. Calibrated risks more closely track empirical defect frequency, especially in dimensions with weaker observability.}
\label{fig:appendix_reliability}
\end{figure*}

In addition to ECE, we visualize the relationship between predicted risk and empirical defect frequency using reliability diagrams. Figure~\ref{fig:appendix_reliability} compares direct hard-threshold interpretation with calibrated posterior risk for the four benchmark dimensions.

The calibrated curves more closely follow the diagonal reference line and preserve monotonic separation between low-, medium-, and high-risk regions. The improvement is particularly visible in Workflow Architecture and Tool-Ecosystem Consistency, where benchmark evidence is more heterogeneous. These observations complement the quantitative ECE results reported in Table~\ref{tab:instantiation_calibration}.

\subsection{SENSITIVITY TO THE WEIGHTING OF PROCESS QUALITY AND CONTROL PRESERVATION}

\begin{figure*}[t]
\centering
\includegraphics[width=0.90\textwidth]{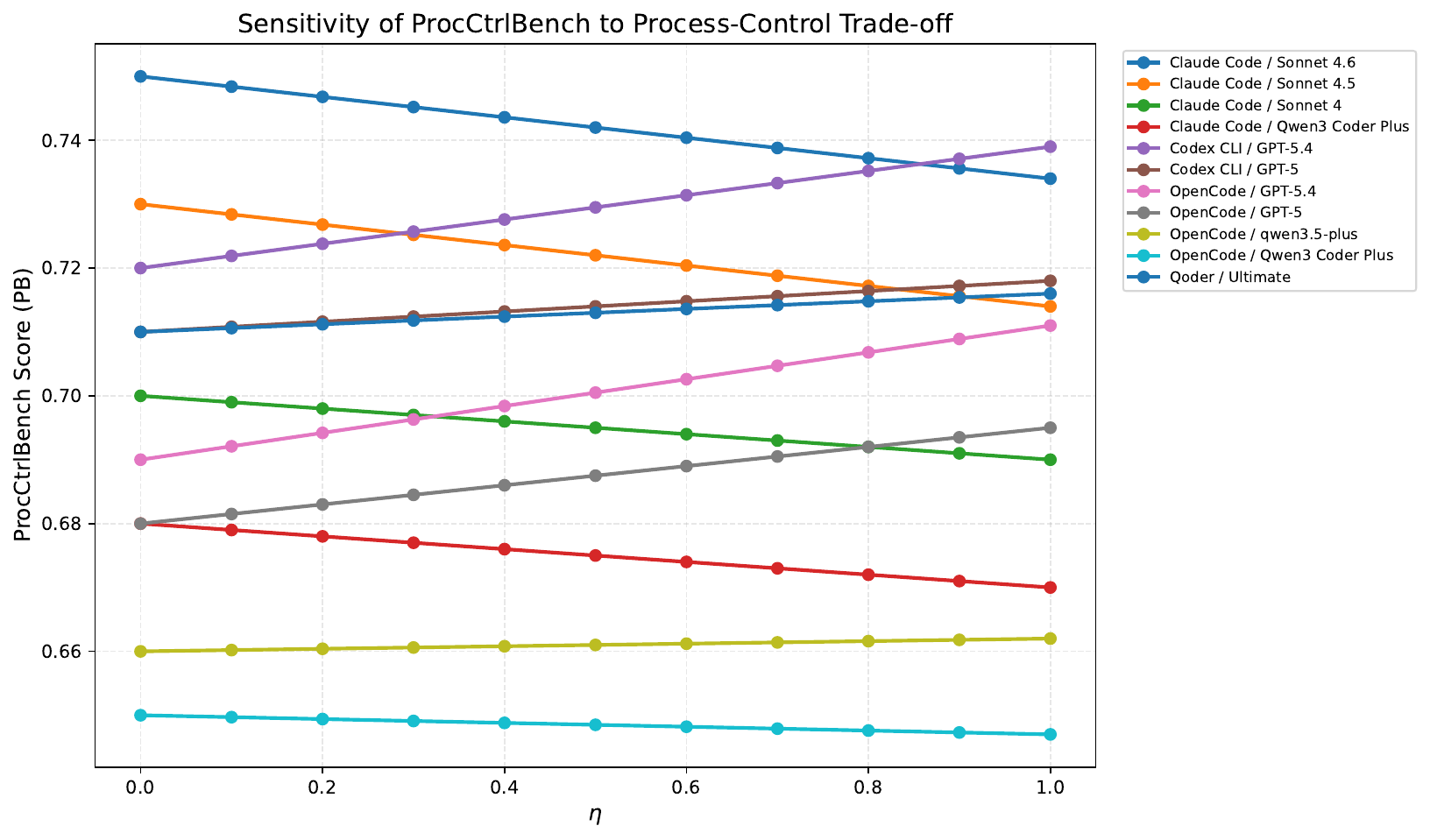}
\caption{Sensitivity of system-level ProcCtrlBench scores to the trade-off parameter \(\eta\) between aggregate process quality and control preservation. Stronger systems remain competitive across a broad range of \(\eta\), while some middle-ranked systems exchange relative order.}
\label{fig:appendix_eta_sensitivity}
\end{figure*}

The overall ProcCtrlBench score combines aggregate process quality and control preservation through the weighting parameter \(\eta\). To examine whether system comparison is overly sensitive to this design choice, we vary \(\eta\) from 0 to 1 and recompute system-level \(PB\) scores. Figure~\ref{fig:appendix_eta_sensitivity} reports the resulting score trajectories.

The strongest systems remain competitive across a broad range of \(\eta\), while some middle-ranked systems exchange relative position depending on the relative weighting of aggregate process quality and control preservation. This result is consistent with our intended use of ProcCtrlBench: \(PB\) provides a useful summary, but the full scorecard remains the primary interpretation surface.

\subsection{Control-Preservation Subdimension Breakdown}

To improve transparency of control-preservation assessment, Table~\ref{tab:appendix_cp_breakdown} reports subdimension-level scores for Interpretability, Interruptibility, Correctability, Reversibility, and Authority Handoff across the 11 evaluated systems.

The breakdown shows that control preservation is not simply the inverse of aggregate defect burden. Some systems maintain relatively strong interpretability and interruptibility despite weaker process efficiency, while others exhibit stronger process quality but less balanced control characteristics. 
This further supports using control preservation as an important component in quantifying execution-process quality, rather than treating it as a simple derivative of process-defect counts.

\begin{table*}[t]
\centering
\caption{Control-preservation subdimension scores across the 11 evaluated systems. Higher is better.}
\label{tab:appendix_cp_breakdown}
\setlength{\tabcolsep}{4.2pt}
\renewcommand{\arraystretch}{1.08}
\begin{threeparttable}
\begin{adjustbox}{max width=\textwidth}
\begin{tabular}{l c c c c c c}
\toprule
System & Interpretability & Interruptibility & Correctability & Reversibility & Authority Handoff & \(CP\) \\
\midrule
Claude Code / Sonnet 4.6        & 0.77 & 0.74 & 0.76 & 0.73 & 0.75 & 0.75 \\
Claude Code / Sonnet 4.5        & 0.75 & 0.72 & 0.74 & 0.71 & 0.73 & 0.73 \\
Claude Code / Sonnet 4          & 0.72 & 0.69 & 0.71 & 0.68 & 0.70 & 0.70 \\
Claude Code / Qwen3 Coder Plus  & 0.70 & 0.66 & 0.68 & 0.67 & 0.68 & 0.68 \\
Codex CLI / GPT-5.4             & 0.74 & 0.72 & 0.73 & 0.70 & 0.72 & 0.72 \\
Codex CLI / GPT-5               & 0.73 & 0.70 & 0.72 & 0.69 & 0.71 & 0.71 \\
OpenCode / GPT-5.4              & 0.70 & 0.68 & 0.69 & 0.67 & 0.69 & 0.69 \\
OpenCode / GPT-5                & 0.69 & 0.67 & 0.68 & 0.66 & 0.68 & 0.68 \\
OpenCode / qwen3.5-plus         & 0.67 & 0.65 & 0.66 & 0.65 & 0.66 & 0.66 \\
OpenCode / Qwen3 Coder Plus     & 0.66 & 0.64 & 0.65 & 0.64 & 0.65 & 0.65 \\
Qoder / Ultimate                & 0.72 & 0.70 & 0.71 & 0.69 & 0.71 & 0.71 \\
\bottomrule
\end{tabular}
\end{adjustbox}
\end{threeparttable}
\end{table*}

\end{document}